\documentclass[aps,prl,twocolumn]{revtex4}
\usepackage{epsfig}
\usepackage{afterpage}
\usepackage{epsfig,amsmath}
\usepackage{color}
\usepackage{amsmath}
\usepackage{amssymb}
\usepackage[latin1]{input enc}
\usepackage{graphics}
\usepackage{graphicx}
\usepackage{varioref}
\usepackage{flafter}
\usepackage{fancybox}
\usepackage{subfigure}
\usepackage{floatflt}
\usepackage{natbib}
\usepackage{psfrag}

\begin{document}

\newcommand{\nn}{\nonumber}
\newcommand{\Vol}{{\rm Vol}}
\newcommand{\sign}{{\rm sign}}
\newcommand{\mmoy}{ \left< m \right> }
\newcommand{\LL}{\textsf{L} }
\newcommand{\Tet}{$T^{*}(\LL)$}
\newcommand{\ff}{\frac}
\newcommand{\modif}[1]{\textbf{\texttt{#1}}}
\title{\textbf{Criterion for universality class independent
 critical fluctuations:\\ Example of the 2D Ising model}}

\author{Maxime Clusel$^1$}\email{maxime.clusel@@ens-lyon.fr}
\author{Jean-Yves Fortin$^2$}\email{fortin@@lpt1.u-strasbg.fr}
\author{Peter C.W. Holdsworth$^1$}\email{peter.holdsworth@@ens-lyon.fr}

\affiliation{ $^1$Laboratoire de Physique, \'Ecole normale
sup\'erieure de Lyon, 46 All\'ee d'Italie, F-69364 Lyon cedex 07, France}

\affiliation{ $^2$Laboratoire de Physique Th\'eorique,
Universit\'e Louis Pasteur, 3 rue de l'Universit\'e, F-67084
Strasbourg cedex, France}

\begin{abstract}
Order parameter fluctuations for the two dimensional Ising model
in the region of the critical temperature are presented. A locus
of temperatures \Tet \ and of magnetic fields $B^{\ast}(\LL)$ are
identified, for which the probability density function is similar
to that for the 2D-XY model in the spin wave approximation. The characteristics of the
fluctuations along these points are largely independent of
universality class. We show that the largest range of fluctuations relative to the
variance of the distribution occurs along these loci of points,
rather than at the critical temperature itself and we discuss this
observation in terms of intermittency. Our motivation is the identification of a
generic form for fluctuations in correlated systems in accordance
with recent experimental and numerical observations. We conclude
that a universality class dependent form for the fluctuations is a
particularity of critical phenomena related to the change in
symmetry at a phase transition.


\end{abstract}
\maketitle

\section{Introduction}
There is ubiquitous interest in systems with extended spatial and
temporal correlations, from all areas of physics. Recently,
observations of fluctuations in global, or spatially averaged
measures of such correlations have provided a possible link
between critical phenomena and non-equilibrium systems. That is,
the probability density function (PDF) for order parameter
fluctuations in the low temperature critical phase of the 2D-XY
model, shown in Figures 1 and 2 below, is very similar to
PDFs~\cite{BHP} for fluctuating spatially, or temporally averaged
quantities in turbulent flow~\cite{LPF,PHL,Careras}, for
electro-convection in nematic liquid crystals~\cite{Jim}, for
numerical models of dissipative
systems~\cite{Farago,Noullez,Peyrard,Ruffo} and of "self-organized
criticality"~\cite{PRL2000}, for fluctuations of river
heights~\cite{Danube,Jensen}, as well as for other equilibrium
systems close to criticality~\cite{PRL2000,ZT,PRL2001,Zheng}.
The simplicity of the 2D-XY model allows a complete understanding of
fluctuation phenomena in this case~\cite{PRE2001,BH2002,Racz2}.
The contrary is true for non-equilibrium systems; the lack of
microscopic theory makes the problems extremely complex.
Phenomenological observations and analogy with better understood
systems can therefore be extremely useful.\\
However, for other critical systems, such generic behaviour seems
only to be observed under restricted conditions. For example, it
has been shown that magnetic fluctuations in the two-dimensional
Ising model, at a temperature, \Tet \ (where $\LL$ is the system
size in units of the lattice constant), below but near the
critical temperature $T_c$, are similar to those of the 2D-XY
model~\cite{PRL2000,ZT,PRL2001,Zheng}. Given the important role
played by universality classes in critical phenomena, this is
rather surprising. In fact, it is well established that critical
fluctuations, as measured at $T_c$~\cite{Binder,Binder2,Bruce},
depend, in general on the universality class of the model, as well
as on the shape~\cite{PRE2001} of the sample and on the boundary
conditions~\cite{Racz}. The apparent similarity of the form of the
fluctuations, over and above the universality class of the models
under consideration therefore seems rather puzzling~\cite{ZT}.\\
Given the generality of the above observations in more complex
systems, it is important to understand this point. In this paper
we address it in detail for the two-dimensional Ising model and in
doing so pose the following questions: Is the similarity in form
quantitative, or only qualitative? Is this PDF really a measure of
critical fluctuations in the problem? Finally can we, from this
investigation shed any further light on the reason for the
apparent "super-universality" observed in the wide range of
experimental and numerical systems ?\\
In answer to these questions, we show that the distribution
functions for the Ising and XY models at \Tet \ are similar, with
the latter representing an excellent fit over almost any
accessible window of measurements for experimental systems.
However they are not the same functions. Numerical evidence
suggests that the difference will remain in the thermodynamic
limit. The origin of the difference is the structure of phase
space associated with the Ising transition. For large amplitude
fluctuations the system, localized in one half of phase space
is able to surmount the barrier separating it from the other,
symmetric half of phase space. If instead one approaches the
critical point by applying a small magnetic field, symmetry
remains broken and one finds a field $B^{\ast}(\LL)$ giving excellent
quantitative agreement between the PDF for the Ising model and the
2D-XY model.\\
We confirm that, despite the small value of the correlation
length, $\xi$~\cite{ZT,Zheng} at \Tet, the system does show
evidence of correlations on all scales up to a length of the order
of the system size. A consequence of this is the development of
coherent structures, the clusters of spins, up to this macroscopic
scale, that dominate the exponential tail of the distribution.\\
From the above, we conclude that the dependence of the PDF on the
universality class comes from the structure of phase space and the
way in which symmetry is restored on passing through the phase
transition. If long range correlations develop,  without the
insuing fluctuations exposing a structured phase space, such as
occurs in the Ising transition, then the form of the fluctutations
will be largely independent of universality class. This is the
case in the 2D-XY model~\cite{ABH} and we propose that it is key
to the approximate "super-universality" observed in a large array
of correlated systems.\\
\section{Order parameter fluctuations in the 2D Ising model}
The Hamiltonian for the Ising model is given by:
$$ {\cal{H}}=-J \sum_{<i,j>} S_i S_j,\; S_i=\pm 1$$
with $J>0$, the exchange constant. We study the model on a square
lattice of size $N=$\LL$\times$\LL, with periodic boundary
conditions and with lattice parameter $a=1$. The order parameter
is defined as the modulus of the magnetization:
$$ m= \left| \ff{1}{N}\sum_{i=1}^{N} S_i \right|.$$
\begin{figure}[!htb]
\begin{center}
\psfrag{sP(m)}{$\sigma\Pi(m)$}
\psfrag{(m-<m>)/s}{$(m-\mmoy)/\sigma$}
\includegraphics[scale=0.33]{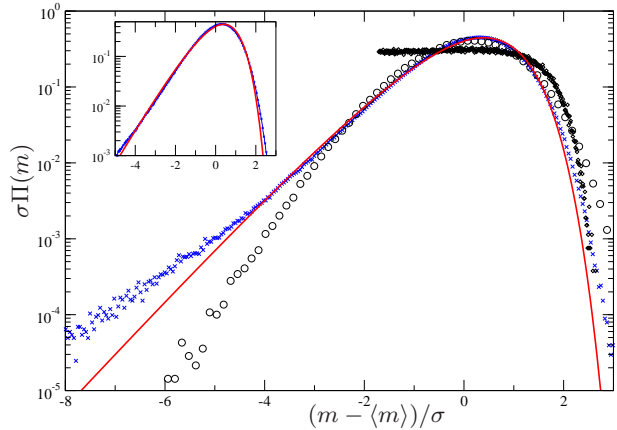}
\caption{Order parameter PDF for $T=2.33J=T_c(\LL=64)$
($\diamond$)\cite{defTc}, $T^*(\LL=64)=2.11J$ ($\times$) and $T=1.54J$ ($\circ$) ,
with $\LL=64$. The inset is a zoom on $\mu \in [-5;3]$.}
\label{L64T}
\end{center}
\end{figure}
\begin{figure}
\begin{center}
\psfrag{sP(m)}{$\sigma\Pi(m)$}
\psfrag{(m-<m>)/s}{$(m-\mmoy)/\sigma$}
\includegraphics[scale=0.33]{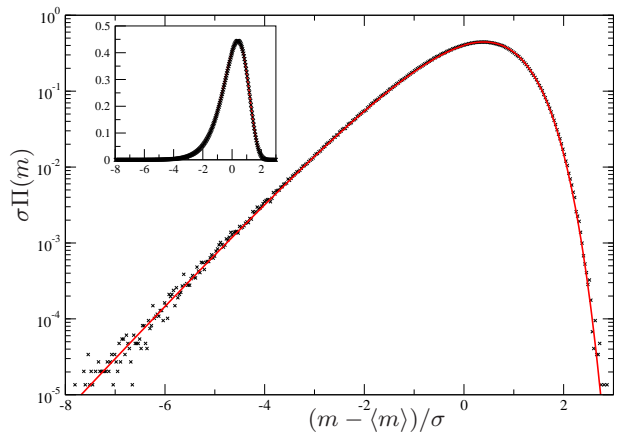}
\caption{PDF for a subsystem \LL=32, in system of size $\LL_0=128$,
for $B=B^*(\LL=32)=0.0035J$ at $T_c(\LL=32)$. The plain curve is the PDF for the
2D-XY model.} \label{bestfit}
\end{center}
\end{figure}
Previous studies, by Bramwell \textit{et al.} \cite{PRL2000},
by Zheng and Trimper~\cite{ZT} and Zheng~\cite{Zheng} suggest that
there is a temperature, \Tet \ just below the critical temperature,
for which the PDF is close to that of the 2D-XY model in the low
temperature phase. It was further suggested that \Tet scales
towards the critical temperature as the thermodynamic limit is
taken and can thus be interpreted as a critical
phenomenon~\cite{PRL2001}. The first step in our study was to
confirm this result. Note that, within the spin wave
approximation, the PDF for the 2D-XY model has the form shown in
Figures \ref{L64T} and \ref{bestfit}, independently of temperature
and hence of critical exponent~\cite{PRE2001}. The PDF was
calculated using the Swendsen-Wang Monte Carlo algorithm
\cite{Swendsen} for various sizes \LL, between 32 and 512. Here we
define \Tet \ as the temperature for which the skewness of the
distribution, $\gamma($\Tet$)$ is equal to that for the 2D-XY
model, $\gamma_{\mathrm{XY}}$, for data restricted to the window
$\mu=(m-\mmoy)/\sigma \in [-6;3]$. We are able to establish this
criterion with a numerical precision $\gamma(T^*(\LL))=\gamma_{\mathrm{XY}} \pm 0.01$,
giving an error for \Tet \ of less than 2\%. This
criterion, though precise, is arbitrary and we could choose
others. However, as it is only an approximate agreement between
the two sets of data, making alternative definitions does not
alter the results discussed below. Figure \ref{L64T} shows
$\log(\sigma\Pi(m))$ against $\mu$ for various temperatures and
for \LL=64. The solid line is the PDF for the 2D-XY
model~\cite{PRE2001}. The best fit is for $T^{\ast}(\LL=64)=
2.11J$, while the critical temperature for the infinite system, as
given by Onsager's exact solution is $T_c=2.25 J$. One can see
that \Tet \ is significantly shifted below $T_c$. At higher
temperatures, the probability for fluctuations below the mean is a
concave function of $|\mu|$ and there is no region approximating
to an exponential tail. The PDF is cut off at a finite value of
$\mu$ corresponding to the constraint $m=0$. As one expects, this
corresponds to a turning point in probability and reflects the
access of the finite size system to the complete
and symmetric phase space.\\
A distribution with an exponential or quasi-exponential tail
cannot correctly describe this minimum and therefore for symmetry
reasons the distribution for the 2D-XY model cannot exactly
describe the data for the Ising model in zero field, for any
temperature below $T_c(\LL)$ \cite{defTc}. However, at \Tet \ the fit is good for
fluctuations out to $-5\sigma$ below the mean
magnetization, corresponding to a probability density
of $10^{-3}$~\cite{Zheng}. From an experimental point of view, reliable data
for $\mu<-6$ would be exceptional \cite{LPF,Jim,Ruffo}, and in this
sense the agreement between the Ising and the XY model data is
very satisfactory (see inset Figure 1). Here we have good statistics
for fluctuations out as far as $-8\sigma$ from the mean, from
which an upturn away from the exponential tail is evident. This is
a consequence of the extremum in PDF at $m=0$. The effect is
independent of system size for the values studied and within the
numerical error obtained, although we cannot exclude the
possibility of corrections to scaling that disappear slowly on the
scale of the sizes studied. For temperatures below \Tet \ the
turning point at $m=0$ is moved outside the accessible window of
measurement, but the PDF is not sufficiently skewed to give a good
fit to the XY model data. For temperatures between \Tet \ and
$T=1.54J$ the the statistics are poor for large fluctuations,
which is consistent with having just a few rare events taking the
magnetization out towards the constraint $m=0$. At lower
temperature still, the distribution crosses over to a Gaussian, as
expected for an uncorrelated system. The relevance of the turning
point in the PDF in the critical region depends on the
universality class and it is one of
the ways in which universality class dependent critical fluctuations appear.

\section{Boundary Conditions and Magnetic Field}
It is clear that the quality of the comparison with the 2D-XY
model would be improved if the turning point in the probability
density at $m=0$ was either displaced or removed. Two obvious ways
in which one might do this are firstly in changing the boundary
condition and secondly in adding a magnetic field. Changing from
periodic to fixed, or window boundaries one can expect to observe
small changes in the form of the universal scaling
function~\cite{PRE2001,Racz}. One might think that these
boundaries would make a finite size system more rigid with respect
to a global spin flip, thus reducing the probability of a
microstate with $m=0$ and improving the quality of the fit. We
have studied the distribution for a window of size $\LL$ embedded
in a larger system of size $\LL_0$. The fits to the PDF for the
2D-XY model are qualitatively better, but the same upturn away
from the exponential tail is observed for fluctuations of more
than about $5\sigma$ from the mean. The situation is not
quantitatively changed compared with periodic boundaries. The data
are not show here.\\
A real quantitative improvement is found however, for the study of
fluctuations in a magnetic field. Approaching the critical point
by fixing $T=T_c(\LL)$ and applying a small field $B$, a field
$B^{\ast}(\LL)$ is defined for which the PDF of magnetic fluctuations
gives the best fit to the 2D-XY data. The data for
$B^{\ast}(\LL)=0.0035J$ for window boundaries, with $\LL=64$ and
$\LL_0=128$, are shown in Figure \ref{bestfit}. To the eye
the quality of the fit is excellent. This is confirmed
quantitatively by measuring the skewness, $\gamma$, and kurtosis
$\kappa$, of the distribution. We find $\gamma= 0.890\pm0.01$,\; $\kappa
= 4.495\pm0.01$, which are in excellent agreement with those of the 2D-XY
model ($\gamma_{\mathrm{XY}}= 0.890$ and $\kappa_{\mathrm{XY}} = 4.489$).
\begin{figure}
\begin{center}
\psfrag{sP(m)}{$\sigma\Pi(m)$}
\psfrag{(m-<m>)/s}{$(m-\mmoy)/\sigma$}
\bigskip
\includegraphics[scale=0.33]{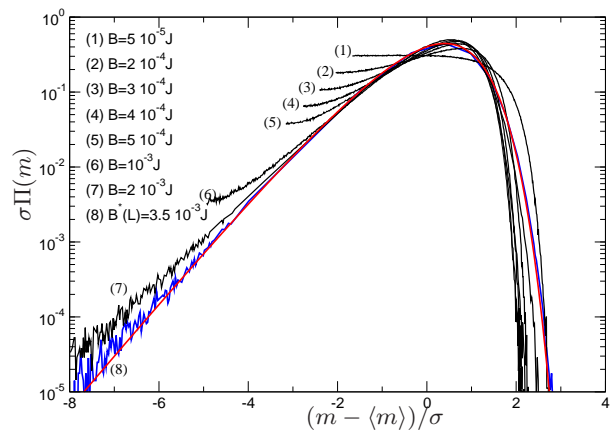}
\caption{PDF for a subsystem \LL=32, in system of size $\LL_0$=128,
for various values of the magnetic field $B$ at $T_c(\LL=128)$. The plain curve is the PDF for the 2D-XY model.}
\label{pdf-evolution}
\end{center}
\end{figure}
The application of a magnetic field breaks the symmetry and
removes the minimum from the PDF for the order parameter. This can
be seen in Figure \ref{pdf-evolution}, where we show the evolution
of the PDF, for a window of size $\LL=32$ in a larger system of
size $\LL_0=128$ for field varying between zero and $B^{\ast}(\LL=32)$.
For zero field the minimum in probability at $m=0$ is clearly
visible and the fluctuations below the mean are cut off by the
constraint $m \geqslant 0$. As the field increases the mean
magnetization increases and the variance, $\sigma$ reduces,
pushing the constraint $m \geqslant 0$ out to larger negative
values of $\mu$. At the same time the end point of the
distribution is no longer a minimum, that is, the PDF terminates
with a finite slope. Within the window of observation, one clearly
sees an exponential tail developing which approaches
asymptotically that of the 2D-XY model. For larger fields the
asymmetry reduces and the PDF crosses over to a Gaussian. The
curve for the 2D-XY model therefore seems to define an envelope
giving the maximum possible asymmetry as the field varies in the
region of the critical point. Behaviour for periodic boundary
conditions is similar. One again sees the development of an
exponential tail with the same slope as the 2D-XY model, but for
the best fit the cut off corresponding to $m=0$ remains within the
window $-8<\mu<3$. The data are not shown here.
\section{Correlation length at \Tet}
\begin{figure}[!htb]
\begin{center}
\psfrag{kbT*/J}{$k_B$\Tet$/J$}
\psfrag{1/L}{$1/\LL$}
\includegraphics[scale=0.33]{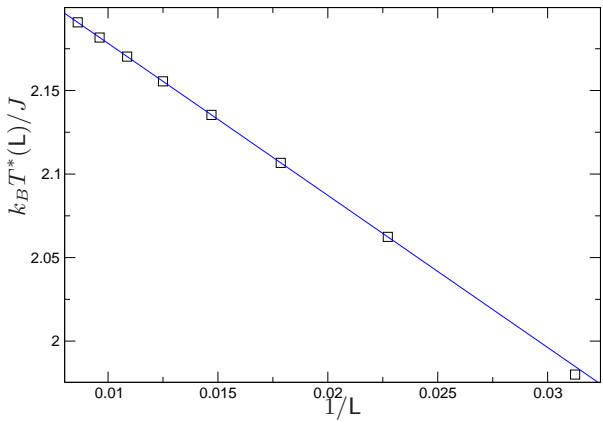}
\caption{Evolution of \Tet \ with \LL. The solid line is the best linear fit.}
\label{TdeL}
\end{center}
\end{figure}

\begin{figure}[!htb]
\begin{center}
\psfrag{G(r)}{$G(r)$}
\psfrag{r}{$r$}
\includegraphics[scale=0.35]{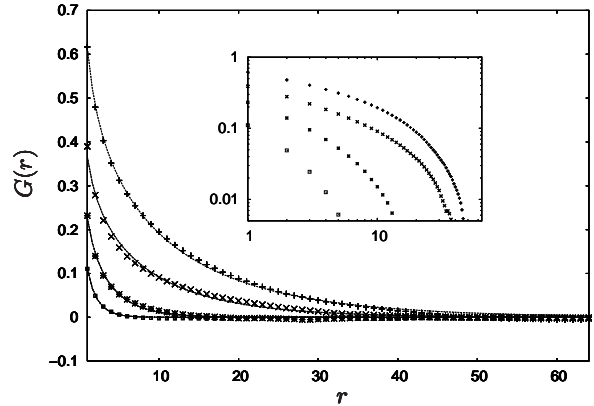}
\caption{Correlation functions for $\LL=128$ and various temperatures $T=2.30J(+),\ 2.27J(\times),\ 2.24J(\ast),\ 2.17J(\square)$. The plain curves are the best fits.}
\label{Autocorr}
\end{center}
\end{figure}

In this section we concentrate on the critical properties of \Tet.
Figure \ref{TdeL} shows how \Tet \ scales with system size. As noted
in references~\cite{PRL2000,ZT}, \Tet \ scales with \LL
as~: $T_c-T^{*}(\LL) \varpropto
\LL^{-1/\nu},$ with $\nu \simeq 1$, the expected value conforming
to the scaling hypothesis for the 2d Ising model.
These finite size scaling results imply that the magnetic
correlation length $\xi(\LL)$ is diverging with the system size at
\Tet, such that $\xi(\LL)/\LL=\mathrm{constant}$. This is
confirmed by the observed universality of $\Pi(\mu)$ along this
locus of points~\cite{Binder}. The correlation length can be computated from
the spin-spin connected correlation function:
$${\rm{G_{\LL}}}(|\boldsymbol{r}_i-\boldsymbol{r}_j|) \: = \:
\left<S(\boldsymbol{r}_i).S(\boldsymbol{r}_j)\right>-\left<S(\boldsymbol{r}_i)\right>.
\left<S(\boldsymbol{r}_j)\right>.$$ Different curves, obtained for \LL=128
and various temperatures are plotted on Figure \ref{Autocorr}.\\
The numerical data is fitted well by the expression
$$ {\rm{G_{\LL}}}(r,{\rm{T}})= \ff{1}{r^{\eta}}\:
e^{-\ff{r}{\xi_{\LL}({\rm{T}})}},$$ with $\eta=0.24 \pm 0.01$, in
good agreement with the theoretical value $\eta=\ff{1}{4}$. Values
of $\xi(\LL)$ and $\xi(\LL)/\LL$ are shown in table \ref{toto}.
\begin{table}
\begin{center}
\begin{tabular}[t]{|c|c|c|c|c|c|c|c|c|}\hline
 $ \LL$           &  32   &  36    &  40    & 44    &  52   &  56   & 64    & 128 \\ \hline
  $\xi(\LL)$      &  0.83 &  1.0   &  1.1   & 1.25  & 1.45  & 1.42  & 1.7   & 2.9     \\
  $\xi(\LL)/ \LL$ & 0.026 &  0.028 &  0.027 & 0.028 & 0.028 & 0.025 & 0.026 & 0.026   \\ \hline
\end{tabular}
\caption{$\xi$ and $\xi/\LL$ for various \LL.\\}
\label{toto}
\end{center}
\end{table}
The main criticism of Zheng and Trimper~\cite{ZT}, that $\xi(\LL)$
is small, is immediately apparent, however the ratio $\xi(\LL)/\LL
\simeq 0.03$ is indeed constant, to a good approximation,
confirming that the correlation length is diverging with system
size, as conjectured. In this respect we are indeed dealing with a
critical phenomena despite the small values measured for the
system sizes considered. This "small, yet diverging" quantity
plays an important role in the approximate universality observed
between disparate systems.

\section{Cluster sizes and distribution }

\begin{figure}[!htp]
\centering
\fbox{\includegraphics[width=7cm, height=7cm]{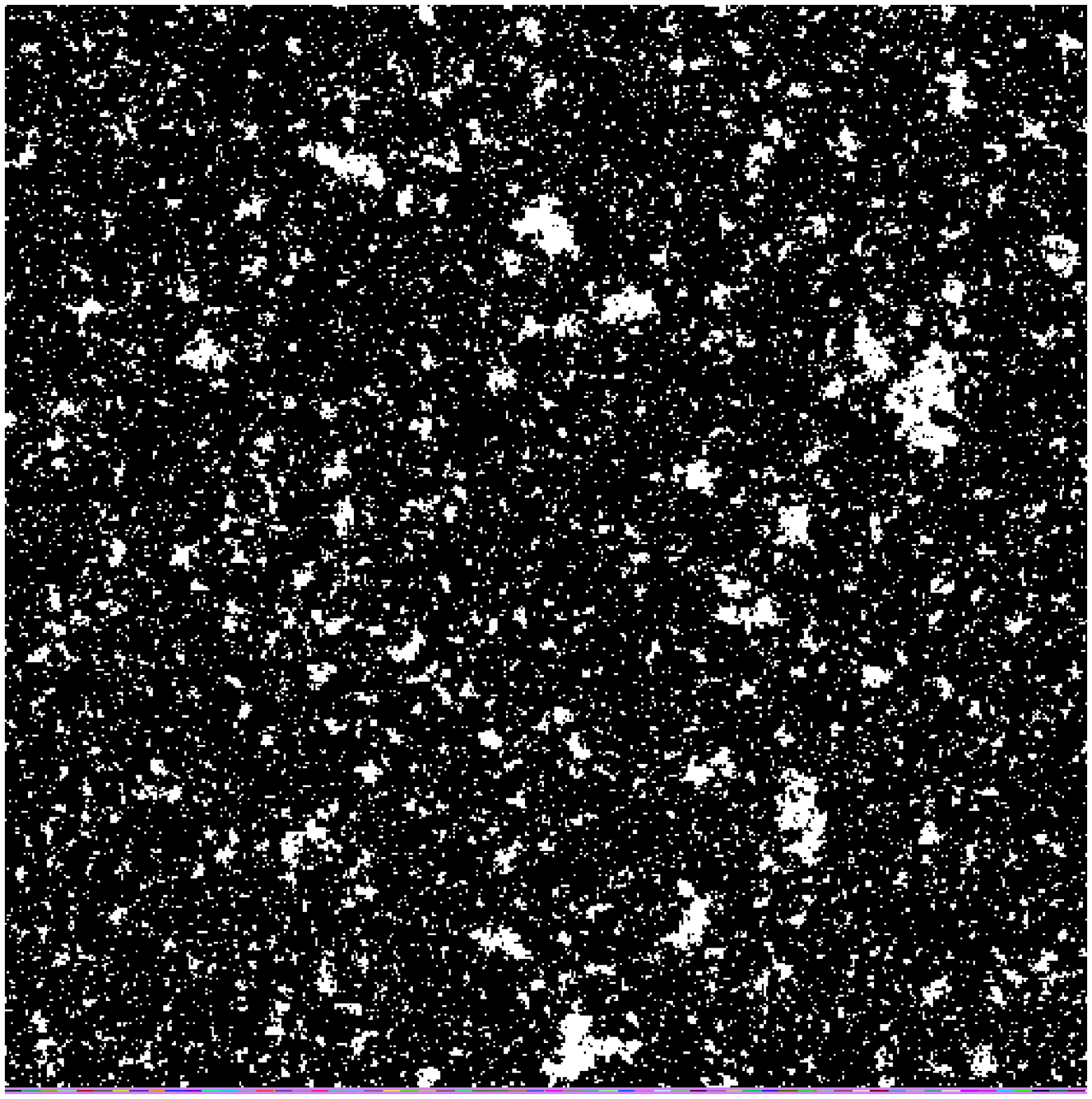}}
\caption{Typical configuration at \Tet.}
\label{conftyp}
\centering
\fbox{\includegraphics[width=7cm, height=7cm]{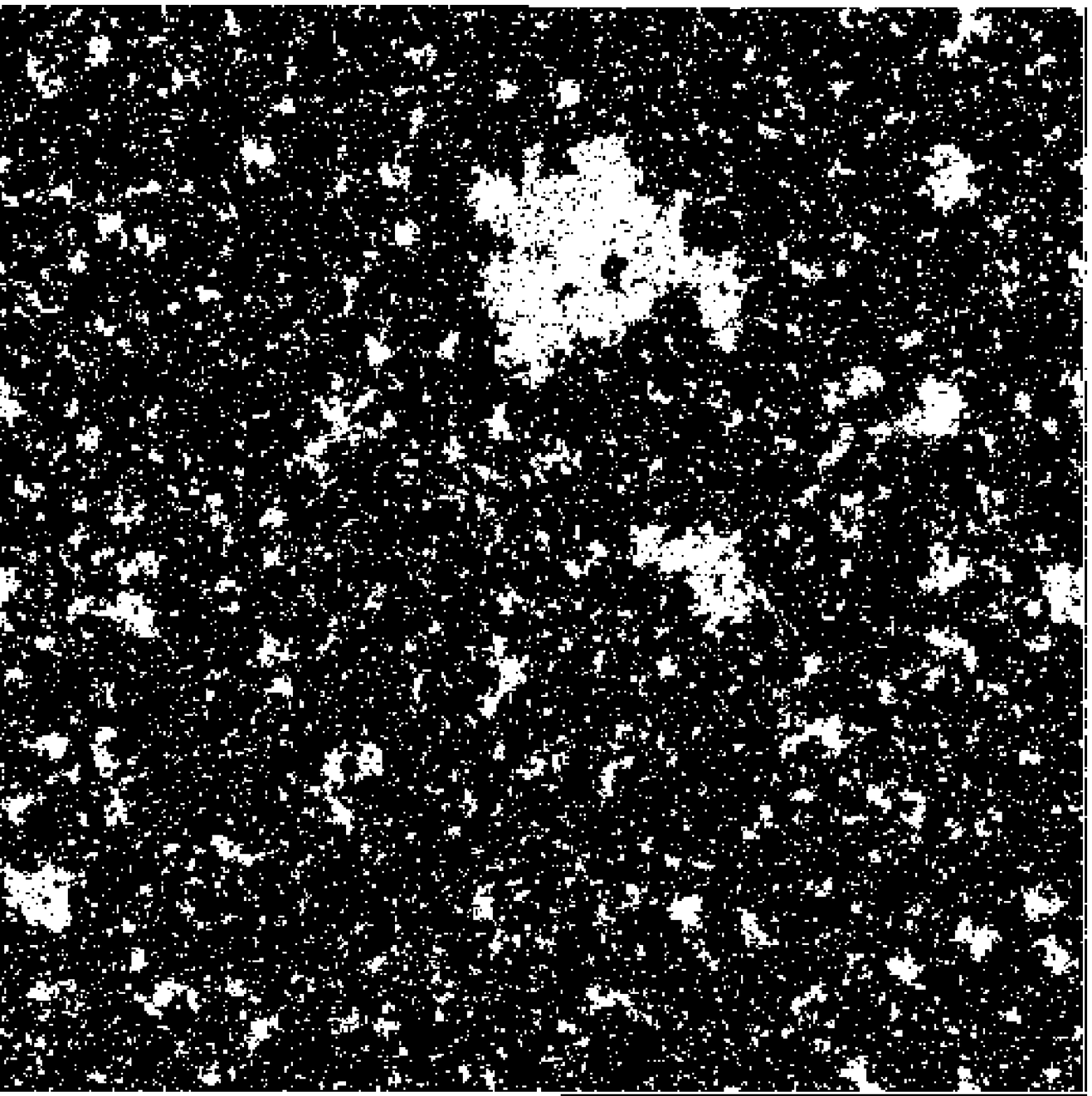}}
\caption{Rare event at \Tet. }
\label{confext}
\centering
\fbox{\includegraphics[width=7cm, height=7cm]{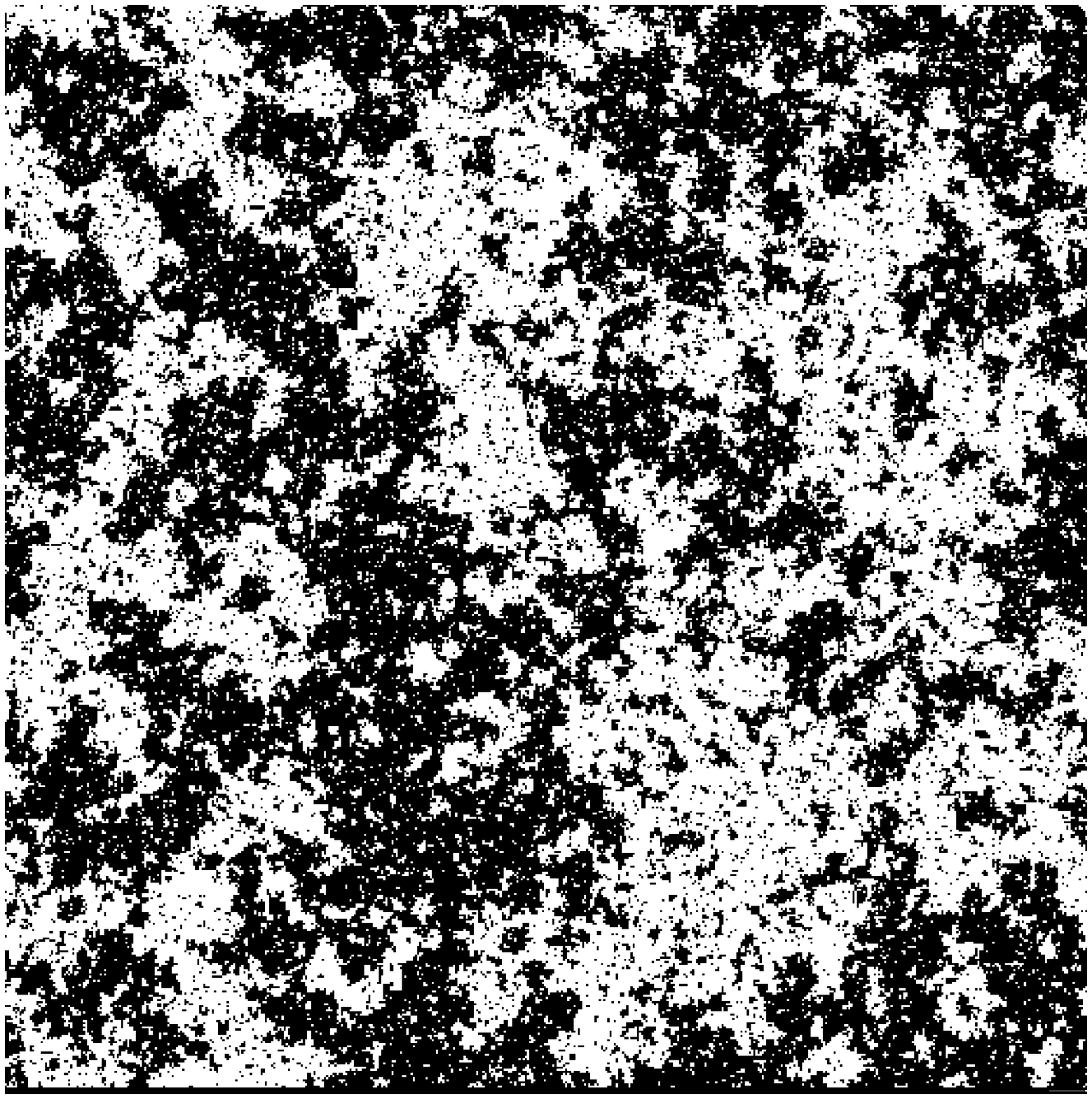}}
\caption{Typical configuration at $T_c$. }
\label{confTc}
\end{figure}

\begin{figure}[!htb]
\begin{center}
\psfrag{s=n/N}{$s$}
\psfrag{P(s)}{$P(s)$}
\includegraphics[scale=0.33]{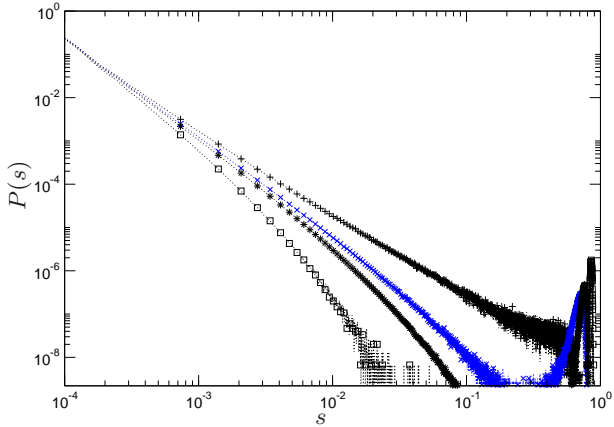}
\caption{Clusters size distribution $P(s)$ for \LL=128 and $T=2.29J(+)$, $T^*(\LL=128)=2.25J(\times)$, $T=2.22J(\ast)$ and $T=2.13J(\square)$ ($s=N_{\mathrm{cluster}}/N$).}
\label{clusterdist}
\end{center}
\end{figure}

\begin{figure}[!htb]
\begin{center}
\psfrag{n/N}{$s$}
\psfrag{P(n/N)}{$P(s)$}
\includegraphics[scale=0.33]{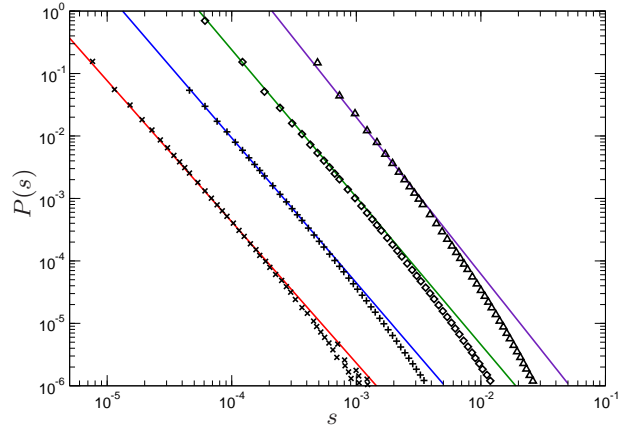}
\caption{Clusters size distribution at \Tet \ for \LL=64($\triangle$), 128($\diamond$), 256($+$) and 512($\times$). The solid lines are the best power laws obtained.}
\label{clusterdist2}
\end{center}
\end{figure}

Definitions of clusters in the Swendsen-Wang algorithm is a
good way to study structures in the 2d Ising model \cite{Hu}. A
cluster is defined as a connected graph of spins in the sense of
reference \cite{Hu}, with  magnetization opposed to the spatially
average value over all the lattice. In concrete terms, a cluster
is a white object in the snapshots shown in Figures \ref{conftyp},
\ref{confext} and \ref{confTc}. Cluster sizes are calculated for
each generated spin configuration and averaged over many
realizations.\\
Given the small value for the correlation length at \Tet \ one might
expect the range of cluster sizes to be extremely limited. One of
the surprises of this study is that this is not the case. One can
get a feeling for this by first studying snap shots. In
Figures \ref{conftyp}, \ref{confext} and \ref{confTc}, we show three
configurations, the first has magnetization $m$ close to the mean
value. One can clearly see a range of cluster sizes up to a
characteristic size that is small compared with $\LL$. The second
snap shot shows a configuration with magnetization four standard
deviations below the mean, $m = \mmoy - 4\sigma$. A much bigger
cluster is present. The large fluctuation is due to the presence
of this large coherent structure. This is very different from what
one would expect for fluctuations of an uncorrelated
system: in this case, a large deviation would correspond to
a configuration with many, small and uncorrelated clusters
appearing spontaneously. This scenario is extremely unlikely,
which is why fluctuations away from the critical point are
Gaussian. Here it is clear that, while the correlation length
fixes the size of typical clusters, much bigger clusters are not
excluded. They are rare events, but not so rare as to be
experimentally irrelevant. This can be compared with the
configuration taken at $T_c(\LL)$, shown in Figure \ref{confTc}.
Here, coherent structures spanning the entire system are not rare.
In this situation the PDF depends strongly on the universality
class and generic behaviour is not expected. Moving along the
locus of points \Tet, the size of typical clusters scales with
$\LL$ through the scaling of $\xi(\LL)$ and we expect the size of
rare clusters to scale in the same way. In this sense the only
length scale in the problem along \Tet \ is $\LL$. Apart from this,
the system is scale free, despite the small values of $\xi(\LL)$
extracted numerically.\\
From the mapping of the Ising model at criticality onto a
percolation problem~\cite{Kastelyn,Hu} one expects the clusters to have a power law
distribution of sizes. Results for the distribution of clusters sizes $P(s)$,
obtained for $\LL=128$ for various temperatures below $T_c(\LL)$
are presented in Figure \ref{clusterdist}.
As the temperature is below $T_c(\LL)$ the system possesses a
spanning cluster whose spin direction defines the global
magnetization direction. Note that, while for standard Metropolis
Monte Carlo the magnetization direction of the spanning cluster
changes extremely rarely, for Swendson-Wang Monte Carlo symmetry
is not broken and the direction oscillates between "spin up" and
"spin down". The spanning cluster is included in the statistics
shown in Figure \ref{clusterdist}. As expected, as the temperature
approaches $T_c(\LL)$ the probability of finding large clusters
increases and the distribution approaches a power law. However,
for temperatures below \Tet \ there is a separation of scales, with
a gap in probability between the largest non-spanning cluster and
the spanning cluster. The gap closes at about \Tet \ and above this
temperature the statistics of the spanning and the secondary
clusters are mixed \cite{JCrem}. When this is the case the largest clusters of
up spins and down spins will be of the same size and there will be
a non-negligible probability of having zero magnetization. The
mixing of statistics of the spanning and secondary clusters is
therefore perfectly consistent with the observation that the
turning point of the PDF, at $m=0$, moves into the window of
numerical or experimental observations for $T$ above \Tet \ and that
the tail of the distribution $\Pi(\mu)$ is no longer well
approximated by an exponential.\\
At \Tet \ the cluster distribution is well represented by a power
law out to cluster size $s=\ff{N_{\textrm{cluster}}}{N} <
\ff{2}{100}$. This is just about the size of the large cluster in
the second snap shot of Figure \ref{confext}. Above this size,
corrections to scaling are manifest as one might expect, but it is
worth noting that the limit of power law behaviour corresponds to
clusters of sizes well in excess of the correlation length
extracted from the autocorrelation function. As shown in Figure
\ref{clusterdist2} the range of application of a power law
distribution increases with system size. For $\LL=512$ it extends
over almost six orders of magnitude of probability and three of
cluster size. The fitted exponent decreases with $\LL$, as shown
in table \ref{alphaL}.
\begin{table}
\begin{center}
\begin{tabular}[t]{|c|cccccc|}\hline
\LL        & 16 & 32 & 64 & 128 & 256 & 512 \\
$\tau$(\LL)   & 3.6 & 3.6 & 2.5 & 2.4 & 2.3 & 2.2 \\ \hline
\end{tabular}
\caption{Exponant $\tau(\LL)$ of the power laws obtained for the cluster size distributions.}
\label{alphaL}
\end{center}
\end{table}
For $\LL =512$ the best fit is for $P(s) \sim s^{-\tau(\LL)}$ with
$\tau(\LL) = 2.2\pm0.1$. This seems to correspond to the
estimated exponent for percolating clusters~\cite{Stauffer,Cambier},
$\tau=2+\ff{\beta}{\nu D_f}\simeq2.1$, where $D_f=187/96$ is the fractal dimension for the 2D Ising model \cite{Stella,Duplantier}, confirmation of the critical
nature of our observations.\\
In this section we have shown that clusters at \Tet \ show, rather
remarkably, both scale free behaviour, and a separation of scales:
Typical cluster size, maximum secondary cluster size and spanning
cluster size are all fixed by $\LL$, however their amplitudes are
sufficiently different to ensure a minimum of interference between
the three scales. In the low temperature phase of the
2D-XY model, the ratio of variance to mean magnetization is,
$\sigma/\left<m\right>=AT/J$~\cite{ABH}, where $A\approx 0.04$. This is a
critical phenomenon, in that the ratio is system size independent.
However, it goes to zero at zero temperature, meaning that the
critical fluctuations have zero amplitude and despite their
singular nature, the system visits an infinitesimal part of phase
space near the mean value of the magnetization. The separation of
scales for cluster formation in the Ising case is analogous to
this. It corresponds to a situation where the fluctuations are
critical, but where the limits of phase space ($m=1$, $m=0$, or
$m=-1$) are not approached. We propose here that observed generic behaviour 
and universality classes independance must have its origin in this point.

\section{Extreme clusters distributions}
\begin{figure}[!htb]
\begin{center}
\psfrag{sP(muk)}{$\sigma_k \Pi(m_k)$}
\psfrag{muk}{$\mu_k$}
\includegraphics[scale=0.33]{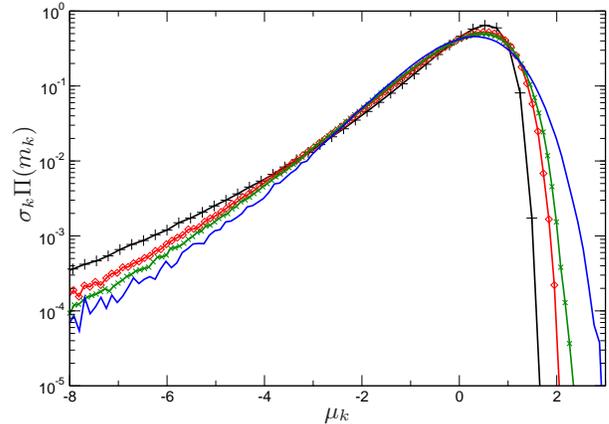}
\caption{PDF of $m_k$, for $k=1(+),5(\diamond),10(\times)$, and non
approximate order parameter (solid line), for $\LL=64$ at $T^*(\LL=64)$.}
\label{Mtilde}
\end{center}
\end{figure}

The form of the distribution in Figure 2 bares a strong resemblance
with Gumbel's first asymptotic solution for extreme
values~\cite{PRL2000,PRE2001}. Indeed there have been a series of
recent papers searching for a connection between this form for
global fluctuation and extreme statistics~\cite{Chap,Racz}. Such a
connection has failed to emerge in Gaussian interface models,
related to the 2D-XY model~\cite{Racz2}, but as there is here
direct access to obvious real space objects, the clusters, it
seems natural to investigate extreme statistics for the clusters. One relevant question is: does the largest cluster dominate $\Pi(\mu)$ rendering the problem of
non-Gaussian fluctuations an extreme value problem for the
clusters ? The answer is no! While the distribution for the
largest cluster is skewed in the right direction and looks
qualitatively quite similar to $\Pi(\mu)$,  it is not sufficient
to reproduce the global fluctuations quantitatively. This
conclusion has been tested in detail: We express the magnetization
as
$$m=1-2\sum_{j=1}^{\infty}\ff{n_j}{\textrm{N}},$$
with $n_j =0, \forall j>j_{\textrm{max}}$. Approximate order
parameters $m_k$ are constructed:
$$m_k =1-2\sum_{j=1}^k\ff{n_j}{\textrm{N}}.$$
If extreme values statistics are relevant for the complete
order parameter PDF, then starting from $k=1$ $\Pi(\mu_k)$, with
$\mu_k =(m_k-\left< m_k \right>)/\sigma$ should converge to
$\Pi(\mu)$ for just a few values of $k$. Results are shown in
Figure \ref{Mtilde} for $k=1,5,10$ and compared with the complete
order parameter PDF. The convergence is slow and $k=10$ is not
sufficient to reproduce well the global PDF. We conclude that all
the clusters are required to reproduce the global fluctuations.
This is not then an extreme value problem, at least in terms of
clusters.\\
This result is rather similar to that obtained for the 2D-XY
model~\cite{BH2002} and for related Gaussian interface
models~\cite{Racz2}, although in the latter more detailed
information on the microscopic distributions is available. The
non-Gaussian fluctuations occur because the largest cluster makes
a macroscopic contribution to the many body sum. However, this
contribution does not dominate, it is the same order as the sum
over all the other clusters. In this sense critical fluctuations
are a marginal case between statistics dominated by the majority,
leading to the central limit, and statistics dominated by a single
event, for example L\'evy statistics.

\section{Intermittency and critical fluctuations}
\begin{figure}[!htp]
\centering
\psfrag{titi}{time}
\psfrag{toto}{$(M-\left<M\right>)/\sigma$}
\includegraphics[width=7cm, height=7cm]{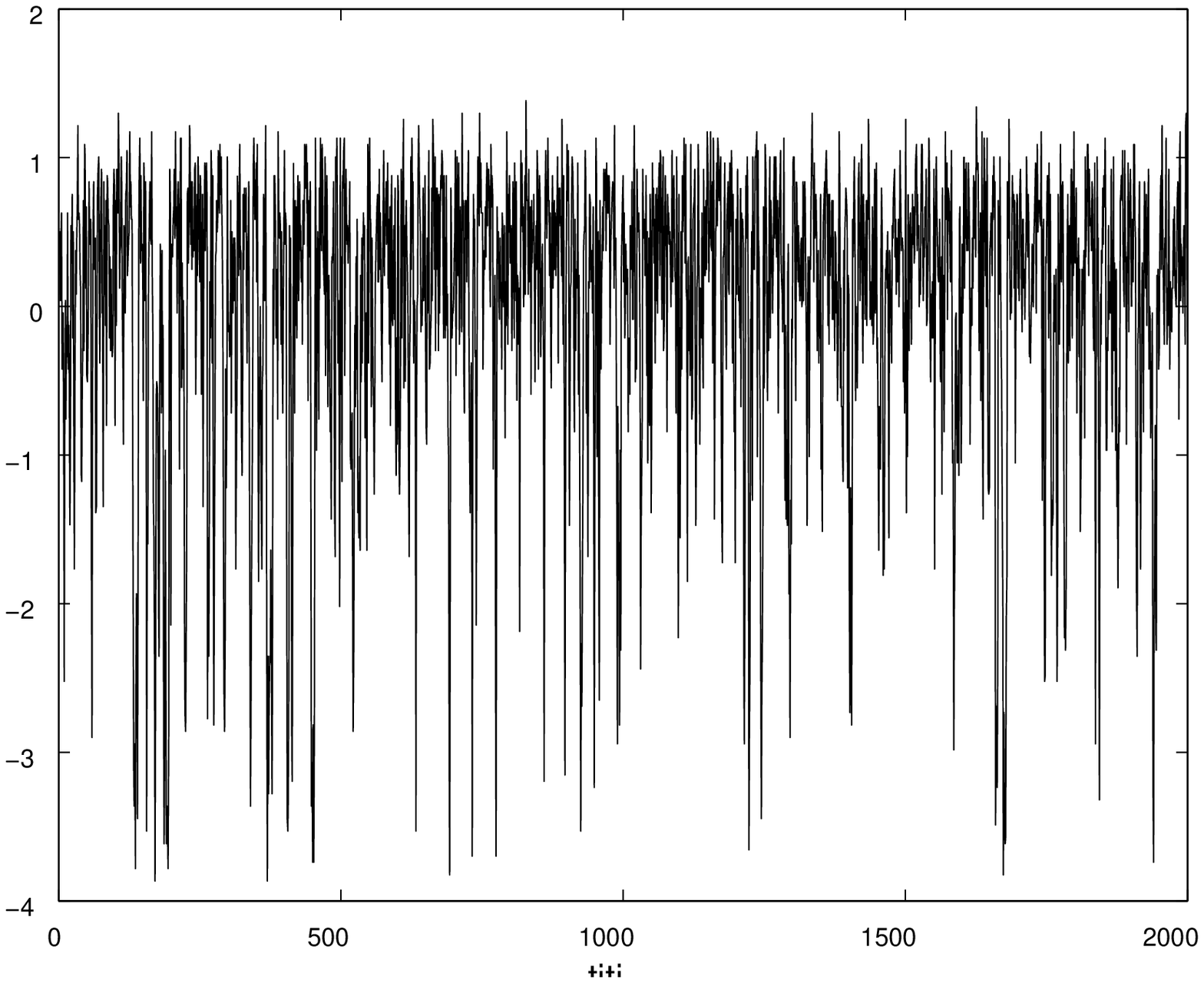}
\caption{Metropolis evolution of $m$ at \Tet.\\}
\label{Intermit1}
\centering
\includegraphics[width=7cm, height=7cm]{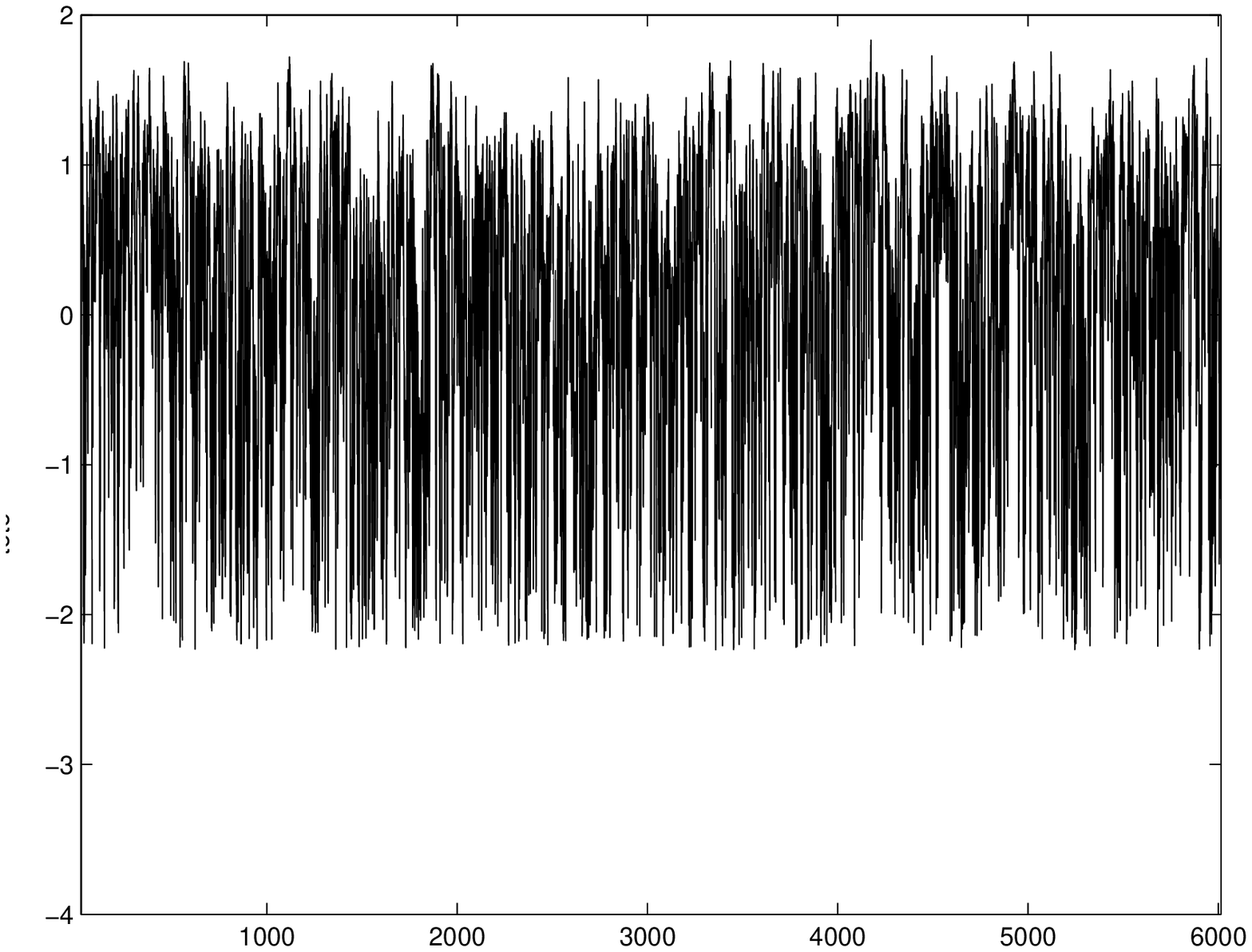}
\caption{Metropolis evolution of $m$ at $T_c(\LL)$.\\ }
\label{Intermit2}
\end{figure}

Figure \ref{Intermit1} shows a sequence of magnetization values for
configurations generated by a Metropolis Monte Carlo algorithm at
temperature \Tet. The asymmetric nature of the distribution is
evident from the stochastic time series. We remark that the data
bare a striking resemblance to the time series of injected power
into a closed turbulent flow at fixed Reynolds number~\cite{PHL}.
The two systems share the characteristic of  making large
deviations from the mean value, on a scale set by the variance of
the distribution $\sigma$. This is the so called "intermittency"
which is so important in studies of turbulence. It is not the
purpose of this paper to discuss the detailed properties of
intermittency, we simply make a comment concerning the scale of
intermittency in the 2D Ising model in the region of the critical
point. In Figure \ref{Intermit2}, a similar time series is shown for
a simulation at $T_c(\LL)$. One can clearly see that the range of
fluctuations, on the scale of $\sigma$ is smaller at $T_c(\LL)$
than at \Tet.
This is an unconventional way of characterizing fluctuations in
the context of critical phenomena. Usually one discusses the
susceptibility, $\chi = (1/T)\sigma^2$, which is a diverging
quantity at a critical point. For a finite size system, $T_c(\LL)$
is generally defined at the temperature where $\chi$ and hence
$\sigma$ is a maximum~\cite{Goldenfeld}.  What we observe here is
that, although the fluctuations on an absolute scale are maximized
at $T_c(\LL)$, on a relative scale, as fixed by $\sigma$ they are
maximized at \Tet. Between \Tet \ and $T_c(\LL)$, large deviations
from the mean value are cut off because of the constraints on the
phase space in which the fluctuations can occur. Specifically, the
free energy barrier between "spin up" configurations and "spin
down" configurations is surmountable with the consequence that the
limits $0< \mmoy <1$ become the relevant limits for fluctuations
and the variance and the mean order parameter values become of the
same size. In this way the magnetic symmetry, which is broken in
the ordered phase, is re-established on passing through the phase
transition. For temperatures up to \Tet, despite the growing
fluctuations the system behaves {\it almost} as if only a single
phase existed, for which the order parameter could vary within the
limits $-\infty < m < \infty$. The distribution for the 2D-XY
model corresponds to the case where this is a limitingly good
approximation \cite{ABH}. If, therefore we make a rather loose
definition of intermittency, as the tendency to make fluctuations
away from the mean, on the scale of $\sigma$ with a probability
that differs from that for a Gaussian function, then maximum
intermittency occurs at \Tet \ and not at $T_c(\LL)$.
\section{Conclusion}
In this article we have compared the order parameter fluctuations
of a finite size 2D Ising model, in the region of the critical
temperature, with those of the 2D-XY in the low temperature phase.
Our motivation for doing this is the observation that fluctuations
of global quantities in a wide range of different correlated
systems are of similar form to the 2D-XY model, or equivalent
Gaussian interface models~\cite{Racz}.  Approaching the critical
temperature $T_c$, either along the zero field axis, or applying a
small field, $B$, at the critical temperature one can identify a
temperature \Tet($B=0$) and a field $B^{\ast}(\LL)$, close to
the critical point for which the order parameter fluctuations are
similar to those of the 2D-XY model. We have established the critical scaling behaviour of the locus of
temperatures \Tet. The correlation length $\xi$ diverges with
system size along the locus of temperatures $T^*(\LL)$~, as one might
expect for critical scaling. However, the ratio $\xi/\LL\sim 0.03$
is a small constant. This is a key point: Critical fluctuations
are usually associated with a phase transition and a change in
phase space symmetry. In the 2D Ising model the symmetry between
the two competing phases is reflected in the form of the PDF,
imposing that $m=0$ corresponds to a turning point in the
probability density. However the 2D-XY model is an exception, in
that there is a continuous line of critical points in the low
temperature phase, but no phase transition or associated change in
symmetry \cite{JKKN}. To an excellent approximation the critical fluctuations
occur in an unconstrained phase space~\cite{ABH}. Hence,
criticality in the Ising model can only resemble that in the 2D-XY
model if the change in symmetry is not apparent. This corresponds
to the condition that the ratio $\xi/\LL$ is small. In the case
of $B^{\ast}(\LL)$, the agreement between the PDFs for order parameter
fluctuations in the two systems is exceptionally good: High
quality numerical data for the Ising model are indistinguishable
from the analytical results for the 2D-XY model. However,
differences are observable for data at \Tet. The agreement is
better in the former case, as the field breaks the symmetry
between the two competing ordered phases, thus eliminating the
turning point, or the minimum value  from the PDF. In this case,
the phase space available for fluctuations strongly resembles that
of the 2D-XY model.\\
Driven, non-equilibrium systems, showing strong correlations, such
as turbulent flow~\cite{LPF,Careras}, resistance
networks~\cite{Ruffo}, self-organized critical
systems~\cite{PRL2000}, or growing interfaces, resemble the 2D-XY
model, in that there is no associated phase transition and no
sudden change of limits that constrain the fluctuations as the
correlations build up. We propose that observation of a
non-Gaussian PDF with finite skewness and an exponential tail, for
fluctuations in a global quantity, is a characteristic of
correlated fluctuations in an effectively unbounded phase space.
The example of the 2D Ising model serves to show that this is not
the case for critical phenomena associated with a
$2^{\mathrm{nd}}$ order phase transition. However, this is a
detail specific to critical phenomena, related to the fact that
the fluctuations become so large that they allow the system to
explore the whole allowed phase space of order parameter values.
From this analysis one could argue that fluctuations at \Tet \ are
not strong. This is true on an absolute scale: For example the
susceptibility, which is a measure of the variance of the
distribution, is small at \Tet \ compared with the maximum value
from which one defines $T_c(\LL)$. However at this temperature
extreme fluctuations compared with the variance are capped by the
constraint, $0 < m < 1$. Parameterizing in terms of the reduced
variable $\mu =(m-\mmoy)/\sigma$ changes this conclusion:
Fluctuations in $\mu$ are essentially unbounded at \Tet \ while they
are constrained at $T_c(\LL)$. Hence the largest fluctuations in
$\mu$ occur at \Tet \  and not at $T_c(\LL)$. It is the variance of
the distribution which defines the scale of the fluctuations that
one might observe experimentally~\cite{LPF} or numerically, and so
in this sense \Tet \ is highly relevant. Physically, this means that
at \Tet \ large fluctuations are rare enough not to modify $\sigma$,
but not too rare to be observed. As long as the constraints
relevant to the phase transition remain unimportant, increasing
the level of fluctuations, that is increasing the ratio $\mmoy /
\sigma$ there will be little, if any evolution of the PDF. This is
exactly what is observed for the 2D-XY model: The PDF has the
universal form discussed above, independently of the critical
exponent along the critical line~\cite{ABH,PRE2001}. This  seems
consistent with the observed generic behaviour in disparate
systems and also appears to be compatible with recent
renormalization group calculations on Gaussian interface models
with quenched disorder~\cite{Krauth}. In the latter a study
is made of the PDF for Gaussian interfaces (of which the low
temperature phase of the 2D-XY model is an example) in the
presence of disorder. This is shown to be highly irrelevant, with
the PDF being unchanged from that of the underlying Gaussian
model.\\
It remains to quantify what we mean by "little  evolution of the
PDF". In this paper we have shown that at \Tet, or
$B^{\ast}(\LL)$, dependence on universality class is largely
absent. It has been further shown that, while the boundary
conditions are important for quantitative comparison, there
effects are not very significant. However, we have remained firmly
in two-dimensions. Moving to three dimensions will no doubt lead
to variations and this will prove an interesting test  for our
explanation of the observed approximate universality for global
fluctuations in correlated systems.
\acknowledgments It is a pleasure to thank S.T. Bramwell for a
critic reading of the manuscript, and  F.Bardou, K. Christensen,
B. Derrida, J. Gleeson, G.Györgyi, H.J. Jensen, J.-F. Pinton, B.
Portelli, A. Metay, Z.R\'acz, J. Richert and C. Winisdoerffer for
stimulating discussions.

\bibliographystyle{apsrev}
\bibliography{biblioCFH}

\end{document}